%% file: make_astro.tex
\documentclass{mpe_report}

\usepackage{psfig,graphicx,epsfig}
\usepackage{color}
\usepackage{amsmath,amssymb,epic,eepic,array}

\begin{document}

\pagenumbering{arabic}
\setcounter{page}{132}

\renewcommand{\FirstPageOfPaper }{132}\renewcommand{\LastPageOfPaper }{133}\include{./mpe_report_Contopoulos}  \clearpage

\end{document}

%% file: mpe_report_Contopoulos.tex
%\documentclass{mpe_report}
%\usepackage{psfig}
%\def\R{~ROSAT}
%\def\RAS{\R all sky survey}
%\begin{document}

\title{Electromagnetic Pulsar Spindown}

\author{Ioannis Contopoulos}  

\institute{Research Center for Astronomy, Academy of Athens,
4 Soranou Efessiou Str., Athens 11527, Greece\\
e-mail: icontop@academyofathens.gr}

\maketitle

\begin{abstract}
We evaluate the result of the recent pioneering 
numerical simulations in Spitkovsky~2006
on the spindown of an oblique relativistic magnetic 
dipole rotator. Our discussion is based on
our experience from two idealized cases, that of an aligned
dipole rotator, and that of an oblique split-monopole rotator.
We conclude that the issue of electromagnetic pulsar
spindown may not have been resolved yet.
\end{abstract}

\section{Introduction}

We believe that
pulsars are spinning down magnetized neutron stars with non-aligned
rotation and magnetic axes (oblique rotators). We also believe that
pulsars lose rotational energy through electromagnetic torques
due to the establishment of large scale electric currents
in their magnetospheres. Unfortunately, an exact expression for
the electromagnetic (Poynting) flux energy loss $L$ as
a function of the magnetic inclination angle $\theta$
remains still elusive.
Most people are content with the approximation
that pulsars spin down at the same rate as
$90^o$ vacuum dipole rotators, namely that
\begin{equation}
L(\theta)\approx L_{\rm vac}(90^o)
\equiv \frac{B^2 r^6 \Omega^4}{6c^3}\ ,
\label{vacuum}
\end{equation}
where, $B$, $r$, $\Omega$ are the polar value
of the magnetic field, the radius of the neutron star,
and the star's angular velocity respectively.
Knowing the exact electromagnetic energy
loss rate, observations of the pulse period $P$ and period derivative
$\dot{P}$ would allow us to accurately infer the value of $B$.
Moreover, the dependence of $L$ on the
magnetic inclination angle $\theta$
has very important implications for the
evolution and distribution of pulsars in the $P-\dot{P}$
diagram (Contopoulos \& Spitkovsky~2006).

In a real pulsar, the vacuum approximation leading
 to eq.~\ref{vacuum} 
is obviously not valid, because without the establishment of
magnetospheric electric currents a magnetic rotator
cannot generate pulses.
The study of the oblique relativistic magnetic rotator
is, therefore, a formidable MHD problem as described clearly in
Spitkovsky~2006: ``Modeling of the structure of the
highly magnetized magnetospheres of neutron stars requires
solving for the self-consistent behavior of plasma in
strong fields, where field energy can dominate the
energy of the plasma. This is difficult to do with the
standard numerical methods of MHD which are forced to
evolve plasma inertial terms even when they
are small compared to the field terms. In these
cases it is possible to reformulate the problem
and instead of solving for the plasma dynamics in strong
fields, solve for the dynamic of fields in the presence
of conducting plasma. This is the approach of
force-free electrodynamics''. 
In his recent pioneering work, Spitkovsky~2006
managed to evolve numerically
an oblique dipolar magnetosphere for about
1.2 turns of the star, and claimed that the solution
very quickly settles to a constant electromagnetic
energy flux which depends on the magnetic inclination
angle as
\begin{equation}
L_{\rm Spitkovsky}(\theta)= 
\frac{B^2 r^6 \Omega^4}{4c^3}(1+\sin^2\theta)\ .
\label{Spitkovsky}
\end{equation}
As we said, this result is very important, and 
while waiting for independent numerical confirmation,
we thought that we should be able to reproduce its main
elements from first principles. We will, therefore, consider
two idealized cases where we know
the electromagnetic energy loss rate analytically.

\section{The axisymmetric magnetic dipole rotator}

Contopoulos, Kazanas \& Fendt~1999 (hereafter CKF) first
obtained the structure of the axisymmetric pulsar
magnetosphere with a dipole stellar magnetic field. 
They showed that, at large distances from the
central star, the asymptotic structure is that of 
a magnetic split monopole, i.e. a certain amount of 
initially dipolar magnetic flux 
\begin{equation}
\Psi_{\rm open}=1.23\frac{\pi Br^3 \Omega}{c}
\end{equation}
stretches out radially
to infinity in one hemisphere, and returns to the star
in the other hemisphere. Electromagnetic
spindown is due to the establishment of a poloidal electric
current distribution 
\begin{equation}
I(\Psi)\sim -\frac{\Omega\Psi}{4\pi}
\left(2-\frac{\Psi}{\Psi_{\rm open}}\right)
\end{equation}
which flows to large distances
along open field lines and returns to the star
through an equatorial current sheet that joins at the
light cylinder with the separatrix between open
and closed field lines. CKF obtained 
a first approximation for $\Psi_{\rm open}$
and $I(\Psi)$ self consistently,
by requiring that the solution be continuous and
smooth at the light cylinder. 
The CKF solution has since been confirmed, improved
and generalized in Gruzinov~2005a, Contopoulos~2005,
Timokhin~2005, Komissarov~2005, McKinnery~2005 and
Spitkovsky~2006.

The electromagnetic spindown luminosity is thus obtained as
\begin{equation}
L_{\rm CKF} = \frac{\Omega}{\pi c}\int_{\Psi=0}^{\Psi_{\rm open}}
I(\Psi){\rm d}\Psi
\approx  \frac{B^2 r^6 \Omega^4}{4c^3}
\approx \frac{\Omega^2}{6\pi^2 c} \Psi_{\rm open}^2
\label{CKF}
\end{equation}
(e.g. Beskin~1997).
This result may be easily understood as follows.
As we said, the structure of the axisymmetric dipole
rotator magnetosphere approaches asymptotically
that of an axisymmetric split-monopole (Michel~1991),
and therefore, the spindown luminosity depends only on 
the amount of open field lines $\Psi_{\rm open}$.

\section{The oblique magnetic split-monopole rotator}

Bogovalov~1999 showed that eq.~\ref{CKF} is also
valid for an oblique split-monopolole rotator.
As long as current sheet
discontinuities are present to guarantee magnetic
flux conservation, properties of cold
MHD plasma flows do not depend on the direction of
the magnetic field. In fact, all properties obtained
for the axisymmetric split-monopole rotator are the same
for the oblique rotator as well.
In particular, rotational losses of the oblique
split-monopole rotator are independent of the
inclination angle. They depend only on the square of
the amount of open field lines
$\Psi_{\rm open}^2$ (obviously, the sign of $\Psi_{\rm open}$
is irrelevant).

We may now derive an important conclusion about the
oblique dipole rotator which too, as we argued,
becomes asymptotically split-monopole-like: rotational
losses of the oblique rotator depend 
indirectly on the inclination angle only through the 
amount of open field lines $\Psi_{\rm open}$ as

\begin{equation}
L_{\rm estimate}(\theta)
\approx \frac{\Omega^2}{6\pi^2 c} \Psi_{\rm open}^{2}(\theta)\ .
\end{equation}

\section{The oblique magnetic dipole rotator}

We have now come full circle. We may estimate
the electromagnetic spindown luminosity of an oblique
rotator through an estimate of the amount of open magnetic
flux as a function of the inclination angle 
$\Psi_{\rm open}(\theta)$. A crude estimate may be obtained
if we calculate the amount
of magnetic flux that crosses the light cylinder distance
in the case of an inclined magnetostatic dipole
at the origin, and rescale our result by the CKF factor 1.23
(note that Gruzinov~2005b used a similar argument in
his estimate of the spindown luminosity). 
This calculation yields
\begin{equation}
\Psi_{\rm open}(\theta)\sim 1.23 \frac{\pi B r^3 \Omega}{c}
(1-0.2\sin^2\theta)\ ,
\end{equation}
which results in the following crude estimate of the
spindown luminosiity,
\begin{equation}
L_{\rm estimate}(\theta)
\sim \frac{B^2 r^6 \Omega^4}{4c^3}(1-0.4\sin^2\theta)\ .
\label{estimate}
\end{equation}
In particular, we see that $L(90^o)<L(0^o)$,
contrary to what both Gruzinov~2005b and Spitkovsky~2006 obtain. 
Note that Spitkovsky's numerical simulation result
agrees well with Gruzinov's estimate, and 
yet the latter mentions that he is not even sure about the sign of
the $\theta$ term that he obtains.

We do not claim here that our above estimate 
has the same weight as the result of a detailed 
numerical simulation.
Interestingly enough, the recent discovery of an ``on/off''
pulsar (Kramer {\em et al.}~2006) offers some further
insight from observations. The pulsar Kramer {\em at al.} 
discovered is thought to be a $90^o$ oblique rotator with
two states of emission. During the ``off'' state, no pulses
are seen, and the neutron star is indirectly inferred to spin down 
at a certain rate $L_{\rm off}$. During the ``on'' state, the
pulsar is observed to spin down at a rate $L_{\rm on}=1.5L_{\rm off}$.
It is very tempting to associate the ``off'' state
with a vacuum state where the spindown rate is analytically
known (eq.~\ref{vacuum}). The observations then imply that
\begin{equation}
L(90^o)=L_{\rm on}=\frac{3}{2}L_{\rm vac}(90^o)
\equiv L_{\rm CKF}
\equiv L(0^o)\ .
\end{equation}
In other words, a $90^o$ oblique MHD rotator is observed to 
spin down at the same rate as an aligned rotator.
This result is more consistent with our crude estimate 
in eq.~\ref{estimate} than with eq.~\ref{Spitkovsky}.

We thus conclude that the discrepancy between the two estimates
is an indication that the issue of electromagnetic pulsar
spindown may not have been resolved yet, and that
further investigation is needed, both analytical and numerical.

%\begin{figure}
%\centerline{\psfig{file=example_figure.ps,width=8.8cm,clip=} }
%\caption{Example of an included figure in a column( .ps file).
%\label{image}}
%\end{figure}

{} 

%\end{document}

%% file: make_astro.bbl
\begin{thebibliography}{} 

\bibitem{B97} Beskin, V. S. 1997, Physics-Uspekhi, 40, 659
\bibitem{B99} Bogovalov, S. 1999, A\& A, 349, 1017
\bibitem{CKF99} Contopoulos, I., Kazanas, D. \& Fendt, C. 1999, ApJ, 511, 351 (CKF)
\bibitem{C05} Contopoulos, I. 2005, A\& A, 442, 579
\bibitem{CK06} Contopoulos, I. \& Spitkovsky, A. 2006, ApJ, 643, 1139
\bibitem{G05a} Gruzinov, A. 2005a, Phys. Rev. Lett., 94, 021101
\bibitem{G05b} Gruzinov, A. 2005b, astro-ph/0502554
\bibitem{McK06} McKinney, J. C. 2006, MNRAS, 368, L30
\bibitem{M91} Michel, F. C. 1991, Theory of Neutron Star Magnetospheres (Chicago: Univ. Chicago Press)
\bibitem{K06} Komissarov, S. S. 2006, MNRAS, 367, 19
\bibitem{Ketal06} Kramer {\em et al.} 2006, Science, 312, 549
\bibitem{S06} Spitkovsky, A. 2006, ApJ, in press (astro-ph/0603147)
\bibitem{T06} Timokhin, A. N., 2006, MNRAS, 368, 1055

\end{thebibliography}
